\documentclass[aps,twocolumns,showpacs, superscriptaddress,nofootinbib]{revtex4}

\usepackage{mathrsfs,pifont,graphicx,amsmath,amssymb,amsthm}
\usepackage[colorlinks=true,urlcolor=blue,linkcolor=blue,citecolor=green]{hyperref}
\newcommand{\beq}{\begin{equation}}
\newcommand{\eeq}{\end{equation}}
\newcommand{\bea}{\begin{eqnarray}}
\newcommand{\eea}{\end{eqnarray}}

\begin{document}

\title{Breaking the $Z_2$ symmetry of the Randall-Sundrum scenario \\
and the fate of the massive modes}

\author{Alejandra Melfo}
\affiliation{Centro de F\'{\i}sica Fundamental, Universidad de Los Andes, M\'{e}rida, Venezuela}
\affiliation{International Center for Theoretical Physics, Trieste, Italy}

 \author{Nelson Pantoja}
 \affiliation{Centro de F\'{\i}sica Fundamental, Universidad de Los Andes, M\'{e}rida, Venezuela}
 
\author{Freddy Ram\'irez}
\affiliation{Centro de F\'{\i}sica Fundamental, Universidad de Los Andes, M\'{e}rida, Venezuela}

\begin{abstract}

We address in detail  the issue of possible resonances in the massive modes on a brane without reflection symmetry.
After identifying a set of solvability conditions, we show explicitly how the modes of the asymmetric case can be 
traced back to the modes of the symmetric RS-2 scenario. The possible occurrence of 
resonances is revisited and discussed by finding analytical solutions. We find that the resonant behavior is very mild even for strong asymmetries, and moreover it occurs only for very large masses, so  that its effects on the Newtonian potential are exponentially suppressed.  

\end{abstract}

\pacs{
04.20.-q, 
11.27.+d  
04.50.+h
}

\maketitle

{\it General considerations.} The set-up for the Randall-Sundrum scenario of
Ref.\cite{Randall:1999vf} (RS-2 scenario) is a single 3-brane 
with positive tension embedded in a $AdS_5$ space with reflection 
symmetry along the extra dimension.  
The problem of a single 3-brane embedded 
in a $AdS_5$ space without reflection symmetry, i.e. with different 
cosmological constants $\Lambda_+$ and $\Lambda_-$ in each side, 
has been considered in 
\cite{Melfo:2002wd,Castillo-Felisola:2004eg, Gabadadze:2006jm, Guerrero:2006gj}. 
In Ref.\cite{Castillo-Felisola:2004eg}, this asymmetric scenario arises 
(rigorously) as the thin wall limit of a self-gravitating thick domain wall 
spacetime generated by a topologically non-trivial scalar field configuration, 
and the Newtonian potential is shown to be the usual one: a dominant four 
dimensional term due to a massless bound state, plus small corrections 
due to the massive modes. Some properties of the 
asymmetric scenarios have been discussed in \cite{Melfo:2002wd,Castillo-Felisola:2004eg, 
Gabadadze:2006jm, Guerrero:2006gj, Padilla:2004tp, Padilla:2004mc, 
Koyama:2005br, Guerrero:2005aw, Bogdanos:2007ti, Bazeia:2007nd, Liu:2009dw}. 
In particular, the occurrence of 
resonances related to the asymmetry has been put forward in \cite{Gabadadze:2006jm}. 

Technically, the evaluation of the contribution of the Kaluza-Klein (KK) modes 
to the Newtonian potential on the brane requires an explicit knowledge of
the graviton wavefunction and, in order to quantize the system,
regulator (negative tension) branes are introduced. In the RS-2 
scenario, due to the assumed reflection symmetry and hence with 
the fifth dimension compactified on an orbifold $S_1/Z_2$, we have 
two branes that represent the boundaries of the fifth dimension. At the end of the
calculation, the regulator brane is taken to infinity and a non-compact fifth dimension 
is thus obtained. The techniques employed 
to obtain the KK modes in the RS-2 symmetric scenario can be extended to the 
asymmetric case without modifications. The evaluation of these 
modes is, however, not as straightforward as in the 
$Z_2$-symmetric case, since the symmetry to characterize this modes is   no longer at our disposal.
Additionally, it may be difficult to fix 
their normalization,  which is crucial 
  to get the correct relative contribution from the zero mode 
as compared to the massive ones in the gravitational Newtonian 
potential on the brane. Given the current interest in asymmetric 
scenarios as brane-worlds, a careful derivation of the massive 
modes is then in order. Since a thorough discussion 
of this problem leads to certain solvability conditions which must be 
satisfied, let us first revisit in some detail the 
evaluation of the KK modes in the RS-2 scenario. 

{\it KK  modes in the symmetric scenario.}
 Let us consider the metric of the RS-2 scenario in conformal
coordinates
\begin{equation}\label{metric}
g_{ab}= e^{2A(z)}\left(\eta_{\mu\nu}\,dx^{\mu}_adx^{\nu}_b +
dz_adz_b\right),
\end{equation}
where $A(z)= -\ln (1+k|z|)$ and
$\eta_{\mu\nu}=\mathrm{diag}(-1,1,1,1)$. In order to find the KK
expansion of the graviton modes we parametrize the graviton
fluctuation in the standard way
\begin{equation}
g_{ab}= e^{2A(z)}\left(
(\eta_{\mu\nu}+h_{\mu\nu})\,dx^{\mu}_adx^{\nu}_b + dz_adz_b\right),
\end{equation}
  and define $h_{\mu\nu}= e^{ip\cdot x}e^{A(z)/2}\psi_{\mu\nu}(z)$, so that
  $\psi_{\mu\nu}$ satisfies the Schr\"{o}dinger
equation
\begin{equation}\label{schoconf-1}
 \left(-\frac{d^2}{dz^2} + V_{QM}\right)\psi_{\mu\nu}(z)=
 m^2\psi_{\mu\nu}(z),\qquad -\infty< z < \infty,
\end{equation}
with
\begin{equation}\label{schoconf-2}
 V_{QM}= 
 \frac{15}{4}\frac{k^2}{(1+k|z|)^2}-3k\delta(z).
\end{equation} 
Integration of (\ref{schoconf-1},\ref{schoconf-2}) across the brane,
$\psi_{\mu\nu}$ being continuous, yields
\begin{equation}\label{jump}
\psi_{\mu\nu}(0^+)=\psi_{\mu\nu}(0^-),\qquad 
\frac{d}{dz}\psi_{\mu\nu}(0^+) - \frac{d}{dz}\psi_{\mu\nu}(0^-)=
-3k\,\psi_{\mu\nu}(0).
\end{equation}
(we shall omit the indices $\mu, \nu$ from now on). For $m^2=0$ the solution of (\ref{schoconf-1}, \ref{schoconf-2}) is well-known to be 
$ \psi_0(z)=  N_0e^{3A(z)/2} $, with $N_0$ a normalization constant.

 Let us focus on the massive modes. How many nontrivial solutions does
(\ref{schoconf-1},\ref{schoconf-2}) have? Let $\varphi(z)=
\psi_m(z)/\psi_m(0)$, with $\psi_m(0)\neq 0$. Since the difference
between two solutions $\varphi_1(z)$ and $\varphi_2(z)$ of
(\ref{schoconf-1},\ref{schoconf-2}) associated to the same
eigenvalue $m^2$ has a continuous first derivative everywhere, this
difference is the {\it classical solution} $\psi_m^{clas}(z)$ of
\begin{equation}\label{schoconf-4}
 \left(-\frac{d^2}{dz^2} +
 \frac{15}{4}\frac{k^2}{(1+k|z|)^2}\right)\psi_m^{clas}(z)= 
 m^2\psi_m^{clas}(z),\qquad \psi_m^{clas}(0)=0,
 \qquad -\infty<z<\infty.
\end{equation}
Following a procedure close to the one employed to obtain Green's
functions for the general self-adjoint problem of the second order
\cite{Stakgold}, the condition $\psi_m^{clas}(0)=0$ can be 
related to the {\it solvability condition} that ensures that the 
(regularized) problem is consistent 
and that every solution of (\ref{schoconf-1},\ref{schoconf-2}) can
then be written as an arbitrary linear combination of $\psi_m^{dist}(z)$
and $\psi_m^{clas}(z)$, where $\psi_m^{dist}(z)$ is any particular 
solution that carries all the singular information (\ref{jump}) and
$\psi_m^{clas}(z)$ is the classical solution of (\ref{schoconf-4}). 
Of course, in (\ref{schoconf-1}), the symmetry 
condition $V_{QM}(z)=V_{QM}(-z)$
has the consequence that for every $m^2$ there exist solutions of
even and odd parity. Since the even solution incorporates the
distributional solution, the odd solution is therefore the classical
one. Being automatically orthogonal to each other, 
\begin{equation}
(\psi^o_m,\psi^e_m)=\int_{-\infty}^{\infty}dz\,\,\psi_m^o(z)^*\,\psi_m^e(z)=0,
\end{equation}
these even (or distributional) and odd (or classical) functions 
appropriately normalized are the two ortonormal modes associated 
to the same eigenvalue $m^2$, which is therefore degenerate.

 The modes should be normalized by requiring
 \begin{equation}\label{dirac-even}
(\psi_{m'},\psi_m)=\int_{-\infty}^{\infty}dz\,\,\psi_{m'}(z)^*\,\psi_m(z)=\delta(m-m').
\end{equation}
However, since the integral in (\ref{dirac-even}) is divergent
$\forall \,m,m'$, some regularization procedure is required.

Following \cite{Randall:1999vf}(see \cite{Callin:2004py} for a detailed 
derivation), we introduce regulator (negative
tension) branes at $\pm z_r$ taking the limit $z_r\rightarrow
\infty$ at the end and the resulting scenario will be called 
the regularized one. Now the gravitational fluctuations
satisfy the additional integrability conditions
\begin{equation}\label{boundary}
\psi_m(z_r^+)=\psi_m(z_r^-),\qquad \frac{d}{dz}\psi_m(z_r^+) -
\frac{d}{dz}\psi_m(z_r^-)= \frac{3}{2}\frac{k}{1 +
kz_r}\psi_m(z_r),
\end{equation}
and these conditions quantize the mass spectrum in units of $\pi/z_r$. Now (\ref{dirac-even}) reads as
\begin{equation}
\int_{-z_r}^{z_r}dz\,\psi_{m_p}(z)^*\psi_{m_q}(z)=\delta_{pq}.
\label{normcond2}
\end{equation}
The corresponding density of states is used to evaluate 
the Newtonian potential, which is then given by 
\cite{Randall:1999vf,Callin:2004py}
\begin{equation}\label{Newton}
V_N(r)=\frac{1}{4\pi M^{3}}\frac{m_{1}m_{2}}{r}\left[|\psi_{0}(0)|^{2}+
\frac{4}{3\pi}\sum_{i=1}^2\int_{0}^{+\infty}dm\,{|\psi_{m}^i(0)|^{2}e^{-mr}}z_{r} \right]\;\;,
\end{equation}
where $M$ is the $5$-dimensional Planck mass.  
 
 Solutions for the even modes are the best known, as they appear in the symmetric case. 
For $m^2\neq 0$ the solution is given by
\begin{equation}\label{even-mode}
\psi_m^e(z)= N_m^e (k^{-1} + |z|)^{1/2}\left\{Y_2[m(k^{-1} +|z|)] -
\frac{Y_1(mk^{-1})}{J_1(mk^{-1})}J_2[m(k^{-1} + |z|)]\right\},
\end{equation}
where $J_n(x)$ and $Y_n(x)$ are the Bessel functions of order 
$n$ of the first and second kind, respectively, and $N_m^e$ is 
a normalization constant determined by (\ref{normcond2}). 
Setting $m_p=m_q=m$ we have
\begin{equation}
(2z_r(N_m^e)^2)^{-1}= \frac{1}{z_r}\int_0^{z_r}dz\,(k^{-1}
+z)\left\{Y_2[m(k^{-1} +z)] -
        \frac{Y_1(mk^{-1})}{J_1(mk^{-1})}J_2[m(k^{-1} +
        z)]\right\}^2
\end{equation}
and we obtain by making use of the asymptotics of the Bessel
functions
\begin{equation}\label{even_normalization}
(N^e_m)^2= \frac{\pi m}{2 z_r}\left[ 1 +
\frac{Y_1^2(mk^{-1})}{J_1^2(mk^{-1})}\right]^{-1}.
\end{equation}

Let us now consider the odd massive modes. For $m^2\neq 0$
the odd solution of (\ref{schoconf-1},\ref{schoconf-2}) is given by
\begin{equation}\label{odd-mode}
\psi_m^o(z)= N^o_m (k^{-1} + z)^{1/2}\left\{Y_2[m(k^{-1} +z)] -
\frac{Y_2(mk^{-1})}{J_2(mk^{-1})}J_2[m(k^{-1} + z)]\right\},\qquad
z>0\,,
\end{equation}
with $\psi_m^o(z)= -\psi_m^o(-z)$, $z<0$, and with $N_m^o$  a normalization constant in the
Dirac's sense of eq.  (\ref{dirac-even}). Since $\psi_m^o(z)$ has a
zero at the brane's position, $\psi_m^o(0)=0$, as follows from
(\ref{jump}) their derivative is continuous at $z=0$ and, as
expected, the odd solutions are unaffected by the brane. 

 We stress that the odd modes have to be normalized by introducing the usual regulator branes. The boundary conditions at $z=\pm z_r$ turn the 
continuous spectrum of masses into a discrete one with 
even and odd modes  sharing the same mass spectrum for 
$z_r\rightarrow\infty$. However, in the symmetric scenario the odd massive 
modes do not contribute to the Newtonian potential at the brane 
located at $z=0$ and  
we obtain the Newtonian potential of Ref.\cite{Randall:1999vf} 
(see also \cite{Callin:2004py}).

{\it  The asymmmetric case.} Next, let us consider the spectrum of gravitational fluctuations of the asymmetric
scenario \cite{Castillo-Felisola:2004eg, Gabadadze:2006jm, Guerrero:2006gj}.
Here, where the gravitational
fluctuations satisfy a Schr\"{o}dinger equation which is not invariant
under $z\leftrightarrow -z$, the massive modes are not functions of
definite parity.  Instead of odd and even modes, we will have weak and distributional ones.  The weak modes will play a key role in 
determining in a consistent way the distributional modes 
that contribute to the gravitation on the brane. Since this point seems 
to be taken lightly on previous works, we will go 
through the calculations in some detail. 

Let $g_{ab}$ be the metric given by (\ref{metric}) with
\begin{equation}\label{Asime0.5}
A(z)=-\Theta(-z)\ln(1-k_{-}z)-\Theta(z)\ln(1+k_{+}z),
\end{equation}
where $k_+$ and $k_-$ are related to the cosmological constants 
$\Lambda_+$ and $\Lambda_-$ at the sides of the brane by
$k_{\pm}=\sqrt{-\Lambda_{\pm}/6}$. It was shown that 
(\ref{metric},\ref{Asime0.5}) can be associated to the metric of an 
asymmetric BPS domain wall spacetime \cite{Castillo-Felisola:2004eg}, 
in the distributional thin wall limit \cite{Guerrero:2002ki}.

Now  $V_{QM}$ is given by
\begin{equation}\label{schoconf-5}
 V_{QM}= \frac{15}{4}\,\frac{k_-^2}{(1-k_-z)^2}\,\Theta(-z) +
         \frac{15}{4}\,\frac{k_+^2}{(1+k_+z)^2}\,\Theta(z) -
         \frac{3}{2}(k_- + k_+)\,\delta(z).
\end{equation}

\begin{figure}
\centerline{
    \includegraphics[width=0.40\textwidth,angle=0]{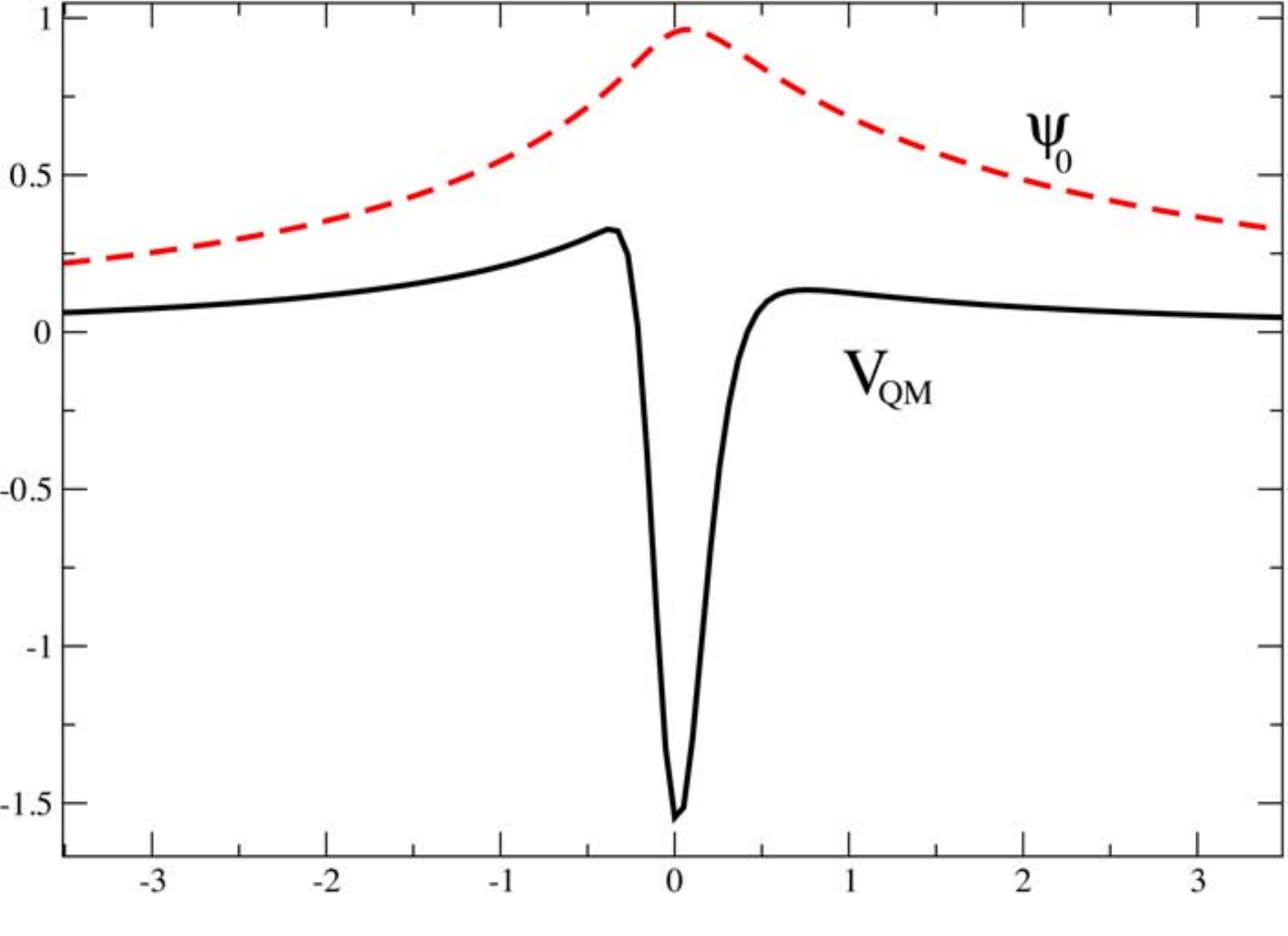}}
\caption{$V_{QM}$ and the (arbitrarily normalized) zero-mode 
\textcolor{red}{$\psi_0$} for $k_->k_+>0$, with a regularized $\delta$.}\label{shape}
\end{figure}

As with (\ref{schoconf-1},\ref{schoconf-2}), the solution of
(\ref{schoconf-1},\ref{schoconf-5}) can be written as an arbitrary
linear combination of a distributional solution $\psi^{dist}_m(z)$, which carries the
singular information
\begin{equation}\label{integrability_condition_1}
\psi^{dist}_m(0^+)=\psi^{dist}_m(0^-)=\psi^{dist}_m(0)\neq 0,
\qquad\frac{d}{dz}\psi^{dist}_m(0^+) - \frac{d}{dz}\psi^{dist}_m(0^-)=
-\frac{3}{2}(k_- + k_+)\psi^{dist}_m(0),
\end{equation}
and a weak solution $\psi^{w}_m(z)$, such that
$\psi^{w}_m(0^+)=\psi^{w}_m(0^-)=\psi^{w}_m(0)= 0$, with a 
continuous first derivative everywhere and therefore 
not affected by the presence of the brane. Although $\psi_m^{w}$ 
is not strictly a classical solution, since 
$\lim_{z\rightarrow 0^+}V_{QM}\neq\lim_{z\rightarrow 0^-}V_{QM}$,
the condition $\psi_m^{w}(0)=0$ can still be related to the solvability 
condition that ensures that the (regularized) problem is consistent \cite{Stakgold}.

For $m^2=0$ there exists a distributional solution 
 $\psi_0(z)=  N_0e^{3A(z)/2}$,
with $A(z)$ given by (\ref{Asime0.5}) and gravity is localized 
on the brane since $\psi_0(z)$ can be normalized, with $N_0$ given by 
\begin{equation}\label{norm_0}
N_0=\sqrt{2}\left[k_-^{-1}+k_+^{-1}\right]^{-\frac{1}{2}}.
\end{equation} 
Fig.  \ref{shape} shows the shape of $V_{QM}$ and the zero mode.

For $m^2\neq 0$, the weak  solution is given by
\begin{equation}\label{weak_1}
 \psi^{w}_m(z)=N^{w}_m\begin{cases}
        +k_+^{-1/2}(k_+^{-1} + z)^{1/2}
        \left[ J_2(mk_+^{-1})Y_2(m(k_+^{-1} + z))-
        Y_2(mk_+^{-1})J_2(m(k_+^{-1} + z))\right],& z > 0 \cr
         & \cr
        -k_-^{-1/2}(k_-^{-1} - z)^{1/2}
        \left[ J_2(mk_-^{-1})Y_2(m(k_-^{-1} - z))-
        Y_2(mk_-^{-1})J_2(m(k_-^{-1} - z))\right],& z < 0\cr
        \end{cases}
\end{equation}
where $N_m^w$ is a normalization constant to be found later.

It should be stressed that the weak solution (\ref{weak_1}) 
is unique, up to a multiplicative constant 
$N_m^w$, while a distributional solution 
is one of an infinite set of particular solutions of 
(\ref{schoconf-1},\ref{schoconf-5},\ref{integrability_condition_1}) 
since, as follows from the solvability condition $\psi_m^{w}(0)=0$, 
any linear combination of (\ref{weak_1}) and a particular solution of 
(\ref{schoconf-1},\ref{schoconf-5},\ref{integrability_condition_1}) 
is also a solution of (\ref{schoconf-1},\ref{schoconf-5},\ref{integrability_condition_1}). 
In absence of $Z_2$ symmetry, $\psi^{w}_m(z)$ and 
a particular $\psi^{dist}_m(z)$, although linearly independent 
solutions, are not automatically orthogonal and can not be identified 
{\it a priori} with the orthonormal massive modes associated to $m^2$, 
a fact that has been overlooked in some previous works 
\cite{Guerrero:2006gj,Bogdanos:2007ti}.
This should not be a problem since in a regularized scenario, 
a Gram-Schmidt process may be used to convert an independent set 
into an orthonormal set with the same span. In the following, the 
regularization procedure of the previous section will be considered.

Introducing regulator branes
 at $\pm z_r$, where the limit $z_r\rightarrow
\infty$ will be taken at the end of the calculations, the gravitational fluctuations
satisfy (\ref{schoconf-1}) but with $V_{QM}(z)$ given by
\begin{eqnarray}\label{schoconf-6}
 V_{QM}&=& \frac{15}{4}\,\Theta(-z)\,\Theta(z_r+z)\frac{k_-^2}{(1-k_-z)^2} +
 \frac{15}{4}\,\Theta(z)\,\Theta(z_r-z)\,\frac{k_+^2}{(1+ k_+z)^2}\nonumber\\
 && - \frac{3}{2}(k_-+k_+)\,\delta(z)+
 \frac{3}{2}\left[\frac{k_-}{1 + k_-z_r}\,\delta(z+z_r) +\frac{k_+}{(1 +k_+z_r)}
 \delta(z-z_r)\right],
\end{eqnarray}
which imposes on $\psi_m(z)$ the integrability conditions at $z=\pm z_r$
\begin{equation}\label{integrability_condition_2}
\psi_m(z_r^+)=\psi_m(z_r^-),\qquad \frac{d}{dz}\psi_m(z_r^+) -
\frac{d}{dz}\psi_m(z_r^-)= \frac{3}{2}\frac{k_+}{1 +
k_+z_r}\psi_m(z_r),
\end{equation}
\begin{equation}\label{integrability_condition_3}
\psi_m(-z_r^+)=\psi_m(-z_r^-),\qquad \frac{d}{dz}\psi_m(-z_r^+) -
\frac{d}{dz}\psi_m(-z_r^-)= \frac{3}{2}\frac{k_-}{1 +
k_-z_r}\psi_m(-z_r).
\end{equation}
As in the previous section, these conditions turn the continuous 
spectrum of massive modes into a discrete spectrum. 

An analogous calculation to that of (\ref{even_normalization}) leads to
\begin{equation}\label{weak_2}
(N_m^w)^2=\frac{\pi m}{z_r}\left[k_-^{-1}\left(Y_2^2(mk_-^{-1}) + J_2^2(mk_-^{-1})\right) + k_+^{-1}\left(Y_2^2(mk_+^{-1}) + J_2^2(mk_+^{-1})\right)\right]^{-1}.
\end{equation}

Next, to obtain the distributional mode $\psi_m^{dist}$ which is 
orthonormal to $\psi_m^w$, we find the solution of 
(\ref{schoconf-1},\ref{schoconf-6}) requiring additionally that
\begin{equation}
(\psi_m^{dist},\psi_m^w)_{z_r}\doteq\lim_{z_r\rightarrow \infty}
\int_{-z_r}^{z_r}dz\, \psi_m^{dist}(z)^*\psi_m^w(z)=0,
\end{equation}
which is evaluated making use of the asymptotics of the Bessel functions.
This provides one and only one solution, up to a multiplicative constant,
which after normalization also in the regularized scenario gives the desired 
orthonormal mode. We find 
\begin{equation}\label{dist_1}
 \psi^{dist}_m(z)=N^{dist}_m\begin{cases}
        (k_-^{-1} - z)^{1/2}\left[ AY_2(m(k_-^{-1} - z))+
        BJ_2(m(k_-^{-1} - z))\right],& z < 0\cr
        & \cr
        (k_+^{-1}+ z)^{1/2}\left[ CY_2(m(k_+^{-1}+ z))+
        DJ_2(m(k_+^{-1}+z))\right],& z > 0 \cr
        \end{cases}
\end{equation}
where
\begin{equation}\label{dist_2}\left.\begin{array}{c}
A=+{k_-^{\frac{1}{2}}\left[J_1^-\left[({Y_2^+})^2+({J_2^+})^2\right]+Y_2^-\left[J_1^+Y_2^+-J_2^+Y_1^+\right]+J_2^-\left[J_1^+J_2^++Y_1^+Y_2^+\right]\right]}\;,\\
B=-{k_-^{\frac{1}{2}}\left[Y_1^-\left[({Y_2^+})^2+({J_2^+})^2\right]+J_2^-\left[J_2^+Y_1^+-J_1^+Y_2^+\right]+Y_2^-\left[J_1^+J_2^++Y_1^+Y_2^+\right]\right]}\;,\\
C=+{k_+^{\frac{1}{2}}\left[J_1^+\left[({Y_2^-})^2+({J_2^-})^2\right]+Y_2^+\left[J_1^-Y_2^--J_2^-Y_1^-\right]+J_2^+\left[J_1^-J_2^-+Y_1^-Y_2^-\right]\right]}\;,\\
D=-{k_+^{\frac{1}{2}}\left[Y_1^+\left[({Y_2^-})^2+({J_2^-})^2\right]+J_2^+\left[J_2^-Y_1^--J_1^-Y_2^-\right]+Y_2^+\left[J_1^-J_2^-+Y_1^-Y_2^-\right]\right]}\;,
\end{array}\right.\end{equation}
with $Y_n^{\pm}= Y_n(mk_{\pm}^{-1})$, $J_n^{\pm}= J_n(mk_{\pm}^{-1})$ and 
\begin{equation}\label{dist_3}
(N_m^{dist})^2=\frac{\pi m}{z_r}\left[A^2 + B^2 + C^2 + D^2\right]^{-1}.
\end{equation}

\begin{figure}
\centerline{
\includegraphics[width=9cm]{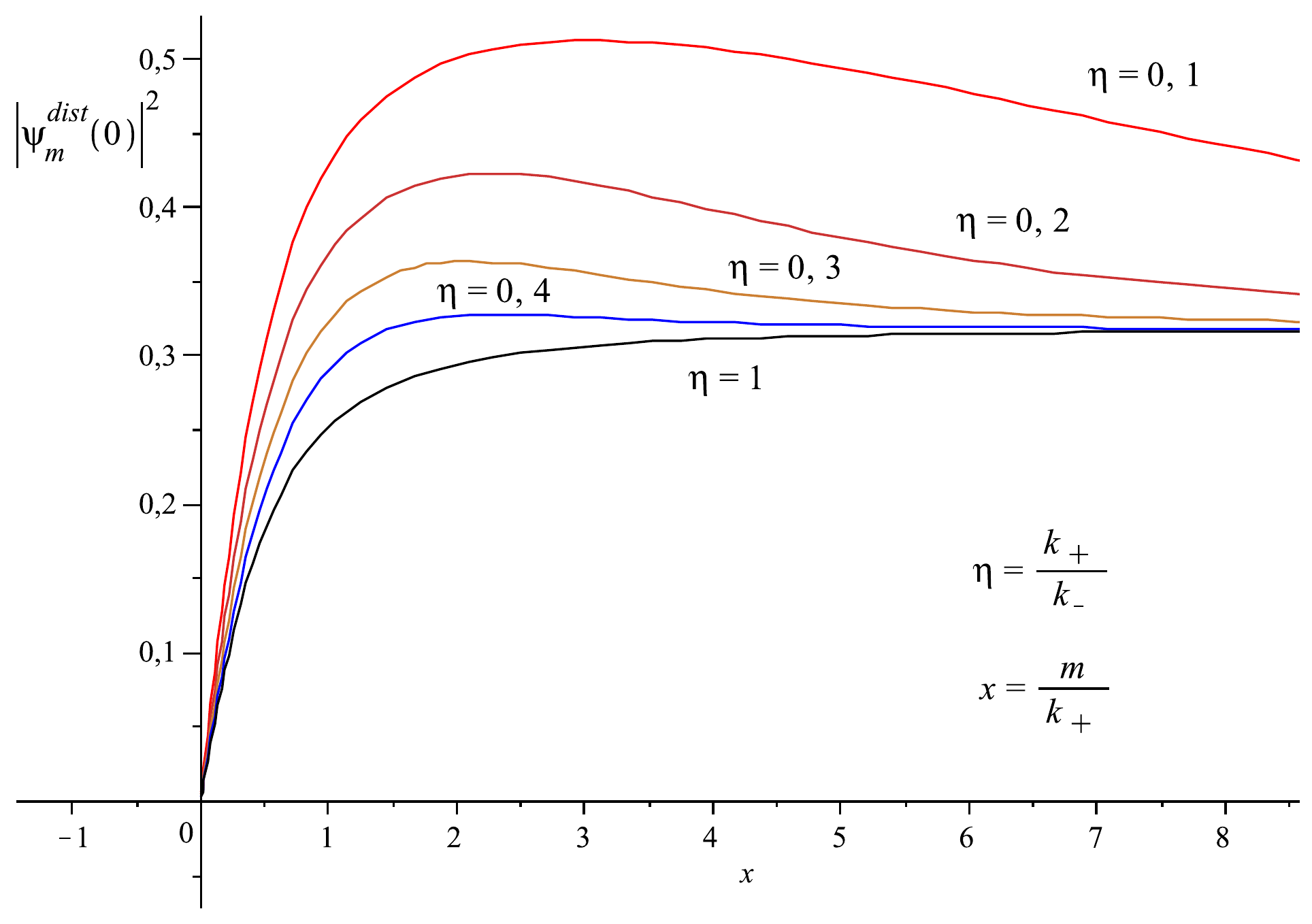} 
\includegraphics[width=9cm]{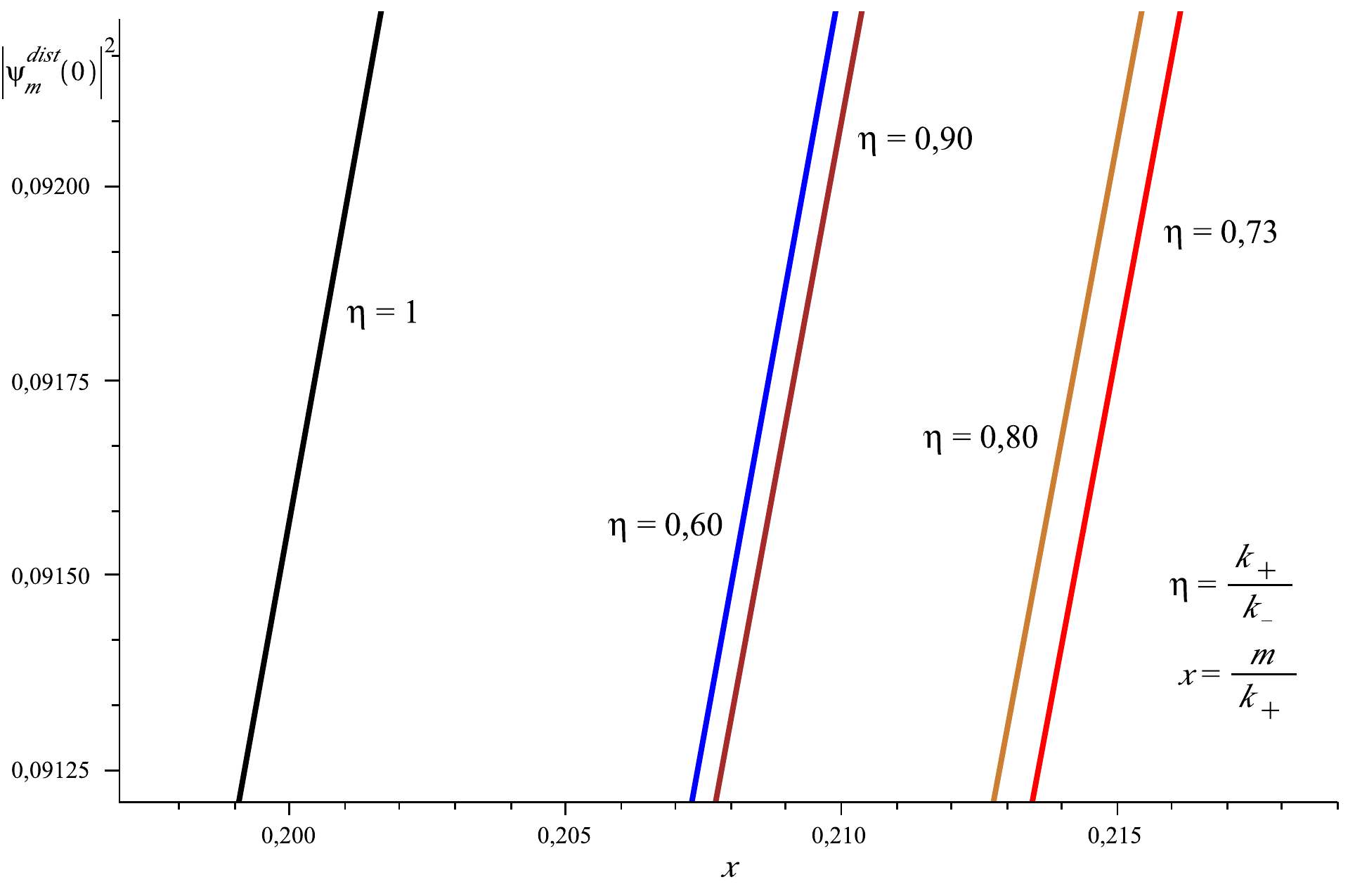}}
\caption{$|\psi_m^{dist}(0)|^2$ for different values of the 
ratio $\eta=k_+/k_-$ as a function of $x=m/k_+$.}\label{massive_shape}
\end{figure}

{\it Resonances.}  We are now ready to discuss   resonances in this scenario. In Fig. \ref{massive_shape}, the value of $|\psi_m^{dist}|^2$ on 
the brane ($z=0$) is plotted for different values of the asymmetry. 
As is expected on general grounds from the shape of 
$V_{QM}$ (see Fig. \ref{shape}), a resonance type behavior 
is observed. Although for strong asymmetries it becomes enhanced, 
this is nevertheless a very mild resonance. For larger values of $k_+$ and $k_-$, it
occurs at very high masses, making  its contribution 
to the Newtonian potential (\ref{Newton}) negligible. Let us define $\eta=k_+/k_- \leq 1$. 
The mass of the resonance, defined as the 
location of the maximum of $|\psi_m^{dist}(0)|^2$, 
its approximately given by $\sqrt{k_-k_+}$ for $\eta\lesssim0.4$. 
On the other hand, as follows 
from  (\ref{norm_0}), the strength of the zero mode on 
the brane decreases for strong asymmetries. 
For $m\ll  k_+ $, from (\ref{Newton}) and 
(\ref{dist_1},\ref{dist_2},\ref{dist_3}), we find that the Newtonian 
potential in the asymmetric scenario is given by
\begin{eqnarray}\label{Newton_asymmetric}
V_N(r)&\simeq&\frac{m_{1}m_{2}}{2\pi M^{3}}\frac{k_{-}k_{+}}{r(k_{-}+k_{+})}\left[1+\frac{2(k_{+}^{2}-k_{-}k_{+}+k_{-}^{2})}{3k_{-}^{2}k_{+}^{2}}\left[\frac{1}{r^{2}}+\left.\frac{6}{r^{4}(k_{-}+k_{+})k_{-}^{2}k_{+}^{2}}\right[k_{-}^{3}\ln{(2k_{+}r)}+k_{+}^{3}\ln{(2k_{-}r)}\right.\right.\nonumber\\ &&\left.\left.\left.-\frac{11}{6}(k_{-}^{3}+k_{+}^{3})+\frac{k_{-}^{5}+k_{+}^{5}}{2(k_{+}^{2}-k_{-}k_{+}+k_{-}^{2})}\right]\right]\right]+\mathcal{O}\left(\frac{1}{r^{7}}\right).
\end{eqnarray}
It follows that we have a four dimensional behavior 
of the Newtonian gravitational potential up to distances $r\sim10^2 \mu m$, 
even for arbitrarily large asymmetries, as far as $\text{min} \{k_-,k_+\} \gg
10^2$ cm$^{-1}$. As expected, for $k_-=k_+=k$, the Newtonian 
potential in the $Z_2$-symmetric RS-2 scenario  
\cite{Randall:1999vf,Callin:2004py} is recovered. 
From (\ref{Newton_asymmetric}), we find that the 
contribution of the massive tower of modes to the term $\sim r^{-3}$ of $V_N$ 
has a minimum for $\eta=\sqrt{3}-1$. Hence, there are slightly asymmetric 
scenarios in which the contribution of the massive modes to $V_N$  
is weaker than in the $Z_2$-symmetric scenario while, 
as follows from FIG. \ref{massive_shape}, for strong asymmetries the 
contribution of these modes grows with the asymmetry.

The occurrence of resonances in the asymmetric scenario, 
as well as the weakness of the zero mode and the strength
of the resonance for strong asymmetries, have been advanced in 
\cite{Gabadadze:2006jm}, where a sharp resonance behavior for 
$m=m_{res}\sim\sqrt{k_-k_+}$ appears in one of two modes, 
being both scattered by the brane. It should be noted that in 
\cite{Gabadadze:2006jm}, a normalization condition is adopted 
which is reduced to the standard one only for $k_-=k_+$. 
From these modes, the gravitational potential on the 
brane is then calculated numerically and compared to that of the 
$Z_2$-symmetric RS-2 scenario with $k^{-1}=(k_+^{-1} + k_+^{-1})/2$, 
finding that they differ the most at scales $r\sim m^{-1}_{res}=
1/\sqrt{k_+k_-}$, and is argued that this result shows that these resonances may 
contribute appreciably to the Newtonian potential on the brane 
\cite{Gabadadze:2006jm}. Since the very same result is obtained 
from (\ref{Newton_asymmetric}), which however receives no 
contribution from masses $m\sim\sqrt{k_+k_-}$, it follows that the 
largest contribution to $V_N$ of the massive modes in the 
asymmetric scenario with respect to the RS-2 symmetric 
scenario should be traced back to the asymmetry, and not to 
the existence of the resonance. 
 
The question naturally arises as to whether (\ref{schoconf-1},\ref{schoconf-6}) 
can have solutions with a clear resonance behavior as in 
\cite{Gabadadze:2006jm}, so we shall 
elaborate a little further on the choice of modes. 
Indeed, any pair $\psi_m^1$, $\psi_m^2$, of orthonormalized 
linear combinations of $\psi_m^{dist}$ and $\psi_m^w$, can be 
taken also as the massive modes in the asymmetric scenario. 
However, in the absence of additional symmetries, any 
other choice of modes different from the orthonormal set 
$\psi_m^{dist}$, $\psi_m^w$, is arbitrary and therefore devoid 
of physical meaning. Let us consider the following example, which shows 
explicitly this arbitrariness. Let $\psi_m^1$, $\psi_m^2$ be given by
\begin{equation}\label{alt_modes}
\psi_m^1(z)=\frac{1}{\sqrt{1 + c^2}}\left(\psi_m^w(z) + 
c\,\psi_m^{dist}(z)\right),\qquad
\psi_m^2(z)=\frac{1}{\sqrt{1 + c^2}}\left(-c\,\psi_m^w(z) + 
\psi_m^{dist}(z)\right),
\end{equation}
where $c$ is a constant which depends arbitrarily on 
$m$, $k_-$ and $k_+$, and $\psi_m^{dist}$, $\psi_m^w$ are given by 
(\ref{dist_1},\ref{dist_2},\ref{dist_3}) and (\ref{weak_1},\ref{weak_2}), 
respectively. The set $\{\psi_m^1,\psi_m^2\}$ is an orthonormal set, 
in the regularized scenario, of solutions of 
(\ref{schoconf-1},\ref{schoconf-6}). Now, we have
\begin{equation}
|\psi_m^1(0)|^2= \frac{c^2}{1+c^2}|\psi_m^{dist}(0)|^2,\qquad
|\psi_m^2(0)|^2= \frac{1}{1+c^2}|\psi_m^{dist}(0)|^2,
\end{equation}
whose shapes depend on $c$. For instance, we can choose the constant 
$c$ such that $c\rightarrow 0$ for $k_-\rightarrow k_+$ and hence 
$\psi_m^1(z)\rightarrow\psi_m^{o}$ and 
$\psi_m^2(z)\rightarrow\psi_m^{e}$ as $k_-\rightarrow k_+$, 
where $\psi_m^o$ and $\psi_m^e$ are the odd and even modes 
of the $Z_2$ symmetric scenario. In any event, a sharp resonance type behavior 
in one of these modes is an artifact introduced 
by an, up to some extent, arbitrary descomposition as (\ref{alt_modes}). 
Nevertheless, since
\begin{equation}
|\psi_m^1(0)|^2 + |\psi_m^2(0)|^2= |\psi_m^{dist}(0)|^2,
\end{equation}
this decomposition gives exactly the same contribution to the 
Newtonian potential on the brane (\ref{Newton}) as the original set 
$\{\psi_m^{dist},\psi_m^w\}$.

{\it Discussion.} We have shown that the calculation of the Newtonian potential arising in asymmetric RS-2 scenarios requires a careful identification of the orthonormal massive modes associated with each value of $m^2$. By normalizing these modes in the standard way, we have revisited the calculations of \cite{Gabadadze:2006jm}. Our analytical solutions show  that the resonant behavior is indeed present, but that it is extremely mild and has no significant contribution to the Newtonian potential. We have shown that the main effect in the Newtonian potential   arises not from the resonances, but  from the  asymmetry itself. Hence, for a wide range of asymmetries, the asymmetric scenario is essentially on the same footing as the original symmetrical one, in terms of the effective 4-dimensional gravitational potential on the brane.


\end{document}